\documentclass[11pt]{article}
\newcommand{\Comment}[1]{{}}
\usepackage{amsfonts,amsthm,amsmath,amssymb,slashed}
\usepackage[textwidth = 430 pt, textheight = 630 pt]{geometry}
\usepackage{color}

\Comment{\usepackage{color}
\definecolor{MyDarkBlue}{rgb}{0.15,0.15,0.45}
\usepackage[linktocpage=true]{hyperref}
\hypersetup{
colorlinks=true,
citecolor=MyDarkBlue,
linkcolor=MyDarkBlue,
urlcolor=MyDarkBlue,
pdfauthor={Jeff Murugan and Horatiu Nastase},
pdftitle={The title},
pdfsubject={hep-th}
}

\usepackage[numbers,sort&compress]{natbib}
\usepackage{hypernat}}
\usepackage{graphicx}
\usepackage{cite}

\newcommand\ignore[1]{}
\def\one{{\,\hbox{1\kern-.8mm l}}}

\def\a{\alpha}\def\b{\beta}

\def\d{\partial}

\def\dslash{\partial\!\!\!/}

\newcommand{\Cset}{{\,\,{{{^{_{\pmb{\mid}}}}\kern-.45em{\mathrm C}}}}}

\newcommand{\be}{\begin{equation}}
\newcommand{\bea}{\begin{eqnarray}}

\newcommand{\ee}{\end{equation}}
\newcommand{\eea}{\end{eqnarray}}

\begin{document}

\renewcommand{\thefootnote}{\fnsymbol{footnote}}

\makeatletter
\@addtoreset{equation}{section}
\makeatother
\renewcommand{\theequation}{\thesection.\arabic{equation}}

\rightline{}
\rightline{}




\begin{center}
{\LARGE \bf{\sc One-dimensional bosonization and the SYK model }}
\end{center} 
 \vspace{1truecm}
\thispagestyle{empty} \centerline{
{\large \bf {\sc Jeff Murugan${}^{a,b}$}}\footnote{E-mail address: \Comment{\href{mailto:jeff.murugan@uct.ac.za}}{\tt 
jeff.murugan@uct.ac.za}}
{\bf{\sc and}}
{\large \bf {\sc Horatiu Nastase${}^{c}$}}\footnote{E-mail address: \Comment{\href{mailto:horatiu.nastase@unesp.br}}{\tt horatiu.nastase@unesp.br}}
                                                        }

\vspace{.5cm}


\centerline{{\it ${}^a$The Laboratory for Quantum Gravity \& Strings, }} 
\centerline{{\it Department of Mathematics and Applied Mathematics, }} 
\centerline{{\it University of Cape Town, Rondebosch, 7700, South Africa}} 
\vspace{.3cm}
\centerline{{\it ${}^{b}$Kavli Institute for Theoretical Physics, University of California, Santa Barbara, CA 93106}}
\vspace{.3cm}
\centerline{{\it ${}^c$Instituto de F\'{i}sica Te\'{o}rica, UNESP-Universidade Estadual Paulista}} 
\centerline{{\it R. Dr. Bento T. Ferraz 271, Bl. II, Sao Paulo 01140-070, SP, Brazil}}

\vspace{1truecm}

\thispagestyle{empty}

\centerline{\sc Abstract}

\vspace{.4truecm}

\begin{center}
\begin{minipage}[c]{380pt}
{\noindent We explore the possibility of extending the familiar bosonization of two dimensions to $(0+1)$- dimensional systems 
with a large number of degrees of freedom. As an application of this technique, we consider a class of SYK-type models, and argue 
that the corresponding action on the gravity dual of the SYK model is given by an STS sequence of dualities.
}
\end{minipage}
\end{center}

\vspace{.5cm}

\setcounter{page}{0}
\setcounter{tocdepth}{2}

\newpage

\renewcommand{\thefootnote}{\arabic{footnote}}
\setcounter{footnote}{0}

\linespread{1.1}
\parskip 4pt

\section{Introduction}
From a value-for-effort perspective, bosonization has to be one of the most versatile tools in low-dimensional quantum field theory. Originally discovered 
in the context of particle physics by Coleman \cite{Coleman:1974bu} as a peculiar relation between the massive Thirring model and sine-Gordon theory, 
it was then sharpened and made
precise by Mandelstam through an operator relation that directly maps the fermionic operators of the Thirring model to 
bosonic operators \cite{Mandelstam:1975hb}. By now there is an extensive literature on the technique and since it is well beyond the purview of this 
note to review all the developments, we refer the reader to the excellent collection of original seminal works with commentary in \cite{Stone:1995ys}. 

The physics underlying bosonization is that particle-hole excitations are effectively bosonic and dominate the fermion spectrum.  This is clear in one spatial 
dimension where the one-particle dispersion near the Fermi level is linear and the particle and hole have nearly the same group velocity at low energies and 
can propagate together coherently as a bosonic quasi-particle. Less so in higher dimensions. However, it is also possible to formulate the bosonization 
transformation as a duality, similar in spirit to the T-duality of string theory \cite{Burgess:1993np}. Since there is nothing intrinsically 2-dimensional about 
dualization, this latter formulation allows for a generalization to more than one spatial dimensions \cite{Burgess:1994tm}, although in these cases it is 
only true as a low energy expansion in a fermionic theory with large mass gap. Moreover, apart from the exceptional case of two spatial dimensions, 
in general the resulting bosonic theory is {\it nonlocal}. In $(2+1)$ dimensions however, the same bosonization-as-a-duality formulation leads to the 
statement that a Wilson-Fisher boson coupled to a $U(1)$ Chern-Simons gauge field at level $k = \frac{1}{2l+1}$ ($l=0,1,2,\ldots$) is equivalent to a 
fermion. This fact, together with the well-known bosonic particle-vortex duality of planar scalar electrodynamics \cite{Peskin:1977kp} as well as a 
recently discovered fermionic analog \cite{Son:2015xqa,Metlitski:2015eka}, leads to a remarkable web of non-supersymmetric 3-dimensional 
dualities \cite{Karch:2016sxi,Murugan:2016zal,Seiberg:2016gmd} that has stimulated much interest in $(2+1)$-dimensional gauge theories. 
It was then shown in \cite{Nastase:2017gxr} that the duality, or at least a version for it with a nonzero mass, and by extension the duality 
web, can be derived from a combination of the basic duality  in \cite{Burgess:1993np,Burgess:1994tm} with particle-vortex duality, in the path 
integral formulation of \cite{Murugan:2014sfa}. By combining two bose-fermi dualities, one finds either a bose-bose duality, or a fermi-fermi duality 
that matches  Son's conjecture in \cite{Son:2015xqa}. In the high energy context, the duality web and its supersymmetric avatar, 3D mirror 
symmetry \cite{Kachru:2016rui}, has expanded significantly on our understanding of Chern-Simons gauge theory coupled to matter \cite{Giombi:2011kc}, 
and their realization in type IIA superstring theory \cite{Murugan:2014sfa}. In the condensed matter setting, it has proven no less important in elucidating 
the properties of topological states of matter,  most notably the physics of the lowest Landau level \cite{Son:2015xqa}. 
 
Given these important developments in two and three dimensions, it is natural to ask if analogous statements exist in one dimension {\it i.e.} in 
quantum mechanics? At first sight, the answer might seem to be obviously no. After all, in one spacetime dimension, the single-fermion Hilbert 
space is 2-dimensional, in contrast to that of a boson. There are, however, several reasons to be hopeful:
        \begin{itemize}
           \item Bosonization has an antecedant in the Jordon-Wigner transformation that relates fermions to spin systems or, equivalently, hard-   
           core bosons through
           \begin{eqnarray}
                c^{\dagger}_{i} = a^{\dagger}_{i} e^{-i\pi\sum_{j<i}a_{j}^{\dagger}a_{j}}\,,
           \end{eqnarray} 
           where $c_{i}^{\dagger}$ ($c_{i}$) are fermion creation (annihilation) operators at site $i$ on a spin chain and $a_{i}^{\dagger}$    
           ($a_{i}$) the corresponding hard-core boson operators at the same site.
           \item Quantum Mechanics can, in some sense, be viewed as a $(0+1)$-dimensional quantum field theory \cite{Boozer:2007zz}. 
           Indeed, many key results of higher dimensional quantum field theory like the diagrammatic expansion of correlation functions, the    
           K\"allen–Lehmann spectral representation of two-point functions, and the LSZ reduction formula find natural analogs in $(0+1)$-   
           dimensions. It is not particularly far fetched to ask if more results find similar expression.
          \item Since each 1-dimensional fermion contributes a 2-dimensional Hilbert space, a collection of $N$ fermions will live in a Hilbert    
          space of dimension $2^{N}$. It is not inconceivable then the dimension counting problem that we encounter with a single fermion can be   
          circumvented in the large-$N$ limit.
       \end{itemize}

Our interest in this problem is two-fold. First, given the utility of bosonization in two and three spacetime dimensions, the one-dimensional problem 
certainly merits study in its own right. Second, quantum mechanics has come under an intense spotlight of late due in no small part to the discovery 
of the Sachdev-Ye-Kitaev (SYK) model. Originally conceived of as a model for spin glasses by Sachdev and Ye \cite{Sachdev:1992fk}, it was modified 
and simplified by Kitaev in a series of seminars in 2015\footnote{A. Kitaev, ``{\it A simple model of quantum holography,}" talk 
at KITP, http://online.kitp.ucsb.edu/online/entangled15/kitaev/, Kavli Institute for Theoretical Physics, UC Santa Barbara, U.S.A., 7 April 2015.} (see also \cite{Kitaev:2017awl}). 
These in turn were further elaborated in the seminal work on the subject \cite{Maldacena:2016hyu}, and extended in several 
directions in \cite{Gross:2016kjj, Berkooz:2017efq, Berkooz:2016cvq, Turiaci:2017zwd, Murugan:2017eto}. 
In its current incarnation, the SYK model encodes the quantum mechanics of 
$N$ Majorana fermions interacting at a single point through an all-to-all random $q$-body. Chief among its many remarkable features, the theory:
\begin{itemize}
   \item is {\it maximally chaotic}, in the sense that it saturates the Maldacena-Shenker-Stanford (MSS) bound on the quantum Lyapunov 
   exponent of $\lambda_{L}\leq 2\pi/\beta$.
   \item is {\it solvable at strong coupling} in the sense that all correlators can in principle be computed by summing the Feynman 
   diagrams at large $N$, 
   \item has an {\it emergent conformal symmetry} at low energies, and
   \item within the context of the general AdS/CFT correspondence \cite{Maldacena:1997re} (see the books 
   \cite{Nastase:2015wjb,Ammon:2015wua} for a review), one expects that the SYK model has a holographic gravity dual. Naively, one 
    expects the SYK model to be the CFT$_{1}$ holographically dual to a two dimensional theory of quantum gravity on $AdS_{2}$. One proposal 
   for the holographic dual to the SYK model is Jackiw-Teitelboim (JT) gravity, a $(1+1)$-dimensional theory of a dilaton coupled to 
   gravity \cite{Maldacena:2016hyu}, though perhaps the true holographic dual is 2+1 dimensional, as argued by \cite{Das:2017pif,Das:2017wae}.
   Either way, the JT gravity model arises by a spherical symmetry reduction of pure 3-dimensional 
   gravity with cosmological constant (see, for example \cite{Sarosi:2017ykf} for an excellent pedagogical review).
\end{itemize}
With the goal of a deeper understanding of the physics of SYK-like systems away from strong coupling and large $N$, we set about 
asking two questions in this article. Firstly {\it is there a fermi-bose dual to the SYK model (or something close to the SYK model)?} 
If such a duality exists in one spacetime dimension, we would expect that, like its higher dimensional counterparts, it would be a 
strong-weak coupling duality that would map the SYK model at strong coupling where it is solvable, to a dual weakly-coupled 
bosonic theory. Combined with holography, this would therefore furnish a novel example of a weak-weak duality. Second we 
ask, {\it how would any such duality act on the bulk dual?} If the analogy to particle-vortex duality in the $(2+1)$-dimensional 
ABJM theory is anything to go by, we would anticipate some combination of S- and T-duality to come into play in the bulk. 
What the details of this are, and their implications, are the subject of the rest of this article.

We organize the paper as follows. In section 2, after a brief review of the standard bosonization formulae in $(1+1)$ dimensions, 
we sketch our argument for bosonization in one dimension through a modified dimensional reduction from two dimensions. We 
outline deficiencies in this treatment and then show how to circumvent these by increasing the number of fermions. Section 3 is 
devoted to our primary application of these ideas to the SYK model and its purported bulk dual. We first present the gravity dual, 
and sketch how it can be embedded in string theory, and then show what the proposed action on it is, and how it matches with 
the action on the boundary. We conclude in section 4 with some speculation on future directions.

\section{Bosonization as a duality in 1-dimension}

Our goal in this section is to develop a $1$-dimensional analog of the familiar bosonization map that
exchanges fermions with bosons in $2$-dimensional spacetime \cite{Coleman:1974bu}.
To this end, it will be useful to review the bosonization procedure in $(1+1)$-dimensions. For concreteness, we will 
follow Mandelstam's original operator formulation of the problem \cite{Mandelstam:1975hb}.

\subsection{Review of bosonization duality in 1+1 dimensions}
While there are by now many examples of systems that are dual under bosonization, it will suffice for our
 purposes to  review the ``canonical" duality between the fermionic massive Thirring model, a 2-dimensional massive
 fermion field with vector interactions, with action
 \be
    S_{\rm mT}=\int \!\!dx\, dt \left[-\bar\psi\gamma^\mu\d_\mu\psi-m\bar\psi\psi-\frac{g}{2}
    j_\mu j^\mu\right]\,,
\ee
and the bosonic sine-Gordon model
\be
   S_{\rm sG}=\int \!\!dx\, dt \left[-\frac{1}{2}(\d_\mu\phi)^2+\frac{\a}{\b^2}\left(
   \cos(\b \phi)-1\right)\right]\,.
\ee
To demonstrate the equivalence, we will start with the Thirring model, where $\bar\psi=\psi^\dagger i\gamma^0$ 
and the fermionic current 
$j^\mu=\bar\psi\gamma^\mu \psi$. We will also use the use a 2-dimensional flat-space metric with
components $\eta^{00}=-1,\eta^{11}=+1$, and gamma matrices
\be
\gamma^0=i\sigma^1\;,\;\; \gamma^1=-\sigma_2\;,\;\;\;
\gamma^5=\gamma^0\gamma^1=\sigma_3,
\ee
in terms of which, the two components of the fermionic current are given by 
\bea
j_0&=&\bar\psi \gamma_0\psi=-i\psi^\dagger (\gamma^0)^2\psi=i(\psi_1^\dagger \psi_1+\psi_2^\dagger\psi_2)\,,\nonumber\\
j_1&=&\bar\psi\gamma_1\psi=i\psi^\dagger \gamma^0\gamma^1\psi=i(\psi_1^\dagger\psi_1-\psi_2^\dagger\psi_2)\,.
\eea
On the right-hand side, `1' and `2' refer to indices of the 2-dimensional fermions. Since the duality is fully 
quantum mechanical we should expect infinities, and since all renormalization constants are finite, these divergences
can be eliminated by an appropriate normal ordering scheme. In this case, we define the quantum currents as 
\be
   \widetilde j^\mu=\lim_{y\rightarrow x}\left[|c\mu(x-y)|^\sigma\bar\psi(x)\gamma^\mu
   \psi(y)+F(x-y)\right]\,,\label{qucu}
\ee
where the numerical constant $c$ is not relevant for our computation, but $c\mu$ will play the role of a unit 
of mass used to define dimensionless quantities. The constant $\sigma$ and function $F$ are chosen so that the 
right hand side remains finite as $y\to\infty$. With these definitions in place, the bosonization formulae are simply
\bea\label{corresp}
\psi_1(x)&=& \sqrt{\frac{c\mu}{2\pi}}e^{\frac{\mu}{8\epsilon}}:\exp\left[-\frac{2\pi i}{\b}\int_{-\infty}^{x} \dot\phi(\xi)\,d\xi-\frac{i\b}{2}\phi(x)\right]:\cr
\psi_2(x)&=& -i \sqrt{\frac{c\mu}{2\pi}}e^{\frac{\mu}{8\epsilon}}:\exp\left[-\frac{2\pi i}{\b}\int_{-\infty}^{x} \dot\phi(\xi)\,d\xi+\frac{i\b}{2}\phi(x)\right]:
\eea
Here the integrations are carried out with an (implicit) adiabatic regulator $e^{-\epsilon \xi}$ inserted.  To determine
the power $\sigma$, we proceed by first expanding the product of the two normal ordered operators as 
\bea
\psi^\dagger_{{}^{1}_{2}}(x)\psi_{{}^{1}_{2}}(y)&=&\mp i\frac{2\pi }{(x-y)}|c\mu (x-y)|^{-\frac{\b^2g^2}{(2\pi)^3}}\times\cr
&&\times:\exp\left[-\frac{2\pi i}{\b}\int_x^y dx' \dot\phi(x')
\mp \frac{i\b}{2}[\phi(y)-\phi(x)]+{\cal O}(x-y)^2\right]:\;,
\eea
in the limit that $x-y\rightarrow 0$. Now if we expand the exponent to linear order in $(x-y)$, drop 
the constant factor because of the normal ordering, and use the definition of the renormalized (quantum) currents
(\ref{qucu}), we find that $\sigma=\b^2g^2/(2\pi)^3$ with corresponding quantum current 
\be
   j^\mu=-\frac{i\b}{2\pi}\epsilon^{\mu\nu}\d_\nu \phi\,.
\ee
The key takeaway from this relation is that the {\it fermionic particle} current is mapped to a  {\it bosonic 
topological} current. Squaring this formula, shows that the current-current interaction in the 
fermionic massive Thirring model turns into a canonical kinetic term for the boson,
\be
-\frac{g}{2}j_\mu j^\mu=-\frac{g\b^2}{8\pi^2}(\d_\mu\phi)^2.
\ee
To evaluate the fermionic mass term, we first construct the two relevant fermion products
\bea
  \psi_2^\dagger(x)\psi_1(y)&\simeq & \frac{c\mu}{2\pi}|c\mu(x-y)|^\delta :e^{-i\b \phi}:\nonumber\\
  \psi_1^\dagger(x)\psi_2(y)&\simeq & \frac{c\mu}{2\pi}|c\mu(x-y)|^\delta :e^{i\b \phi}:\;,
\eea
and note that the only nontrivial part of either expression is the power of the $(x-y)$ factor, given by
\be
    \delta =\frac{g}{2\pi}\left(1+\frac{\b^2}{4\pi}\right)\;.
\ee
Next, collecting the common prefactor of the two products, itself a c-number (that depends on $|x-y|$),  
we find that the mass term
\be
m\bar\psi \psi=m\psi^\dagger i\gamma^0\psi=-m(\psi_2^\dagger \psi_1+\psi_1^\dagger \psi_2)
\equiv \frac{\a}{\b^2}:\cos(\b\phi):\;,
\ee
giving the nontrivial boson interaction of the sine-Gordon model. Finally, the kinetic term of the fermions is slightly 
less simply bosonized. However, the fermion equations of motion (written in components) take the generic form
\begin{eqnarray*}
-i(\dot\psi_2+\psi_2')-m\psi_1&=&...\\
-i(\dot\psi_1-\psi_1')-m\psi_2&=&...\;,
\end{eqnarray*}
(on the right-hand side we have interactions)
suggesting that, at the free level at least, $\psi_1$ and $\psi_2$ are right-moving and left-moving respectively 
on the real spatial line, and that we make the identification $\psi_1\leftrightarrow \psi_{R}$ and 
$\psi_2\leftrightarrow \psi_{L}$. Then, by using symmetric derivatives, for example
\begin{eqnarray}
   \psi^{\dagger}_{R}(x)\psi'_{R}(x) = \lim_{\epsilon\to 0} \psi^{\dagger}_{R}(x)\left(
   \frac{\psi_{R}(x+\epsilon) - \psi_{R}(x-\epsilon)}{2\epsilon}\right)\,,
\end{eqnarray}
expanding the bosonic exponentials to quadratic order in $\epsilon$ and dropping any total derivatives and 
$c$-numbers, we can read off the correspondence
\begin{eqnarray}
   \overline{\psi}\partial\!\!\!/\psi \leftrightarrow \frac{1}{2}(\partial\phi)^{2}\,.
\end{eqnarray}
It is important to emphasise that expressions such as \eqref{corresp} are 
{\it not operator identities}. Instead, they are to be understood as statements that any correlators computed with 
the fermionic fields, in the fermionic vacuum and with a particular cutoff are the same as the 
corresponding correlator, computed with the associated bosonic operator, computed in the bosonic vacuum 
and with the same cutoff.

\subsection{Duality in (0+1)-dimensions}
With this in mind, we now seek to mimic, as closely as possible, the previous duality but in one spatial dimension
 ($d=1$). One immediate problem that arises is that in quantum mechanics a given state can be populated by a single 
fermion, whereas any number of bosons can be put into the same state. However, if the total energy is kept arbitrary,
and we consider the {\em field} in coordinate space and not in energy space, this can be circumvented. 
In other words, if we onsider a quantum mechanical system with a large number of fermionic states, it is plausible 
that we could still realize a bose-fermi duality. With this mind, two strategies immediately present themselves: 
\begin{itemize}
   \item  Kaluza-Klein expanding the (1+1)-dimensional bose-fermi duality on a circle, retaining the entire inifinite tower of KK modes, or 
   \item considering a large $N$ limit of $N$ copies of a $(0+1)$-dimensional model.
\end{itemize}    
Next, we observe that in two spacetime dimensions, canonical scalars and fermions have mass dimensions
$[\phi]=0$ and $[\psi]=1/2$ respectively, while the couplings $\beta$ and $g$ are dimensionless. This is why we needed to introduce the scale $\mu$ and write 
$\psi\propto \sqrt{\mu}$ in the bosonization formula. By contrast, in $(0+1)$-dimensional quantum mechanics, 
$[\phi]=-1/2$ and $[\psi]=0$. A sine-Gordon-like bosonic interaction term,
\be
   V_{b} = -\frac{\a}{\b^2}[1-\cos(\b\phi)]\;,
\ee
must therefore have a dimensional coupling. Indeed since $[\b]=1/2$ and $[\a]=2$, we can put $\a=M^2$, and 
let  $\b$ run with the energy scale. On the other hand, a four-fermi type interaction
\be
   V_f=\frac{g}{2}(\psi^\dagger \psi)^2\,
\ee
requires $[g]=1$, but since a fermion in one dimension has only one component, this begs the question: 
{\it how many fermions are needed in order to construct a Thirring-like model?}

\subsection{Kaluza-Klein compactification and dimensional reduction}
As suggested above, one way to construct a (0+1)-dimensional duality is to consider a circle-compactification of 
the (1+1)-dimensional bosonization, and then attempt a dimensional reduction to the zero modes. 
In a sense, the duality for the theory on a large enough circle is to be expected 
as it is essentially 1+1 dimensional and nothing is lost in the procedure. The dimensional reduction, on the other 
hand, is by no means guaranteed to succeed.  Nevertheless, let us persist. 
Expanding the 2-dimensional fermion $\psi$ in KK modes, 
\be
\psi=\sum_n \psi_n e^{inx}\,,
\ee
reduces the spatial part of the kinetic term to 
\be
   -\bar \psi \gamma^1\d_1 \psi =
   n\psi^\dagger_n \sigma_3\psi_n=-in j_{1,n}=n(\psi^\dagger_{1,n}
   \psi_{1,n}-\psi^\dagger_{2,n}\psi_{2,n}).
\ee
On the other hand, expanding a real 2-dimensional boson as
\be
   \phi=\sum_n \phi_n e^{inx}\,,
\ee
yields a spatial part of the kinetic term
\be
   -\int dx (\d_1 \phi)^2=+nm\int dx e^{i(n+m)x}=-n^2\phi_n^2\,,
\ee
with a similar expression for a complex scalar. The fermion mass term similarly decomposes as 
\be
   -m\bar\psi\psi=m(\psi_1^\dagger \psi_2+\psi^\dagger_2\psi_1)
   =m(\psi_{1,n}^\dagger \psi_{2,n}+\psi_{2,n}^\dagger
   \psi_{1,n})\,.
\ee
Evidently then, we seem to need the mixing of at least two fermionic modes in order to obtain the correct 
bosonized theory. We will make this more precise in due course. In addition to a circle, it will also be useful 
to understand the compactification on an interval. Since this is just
a circle moded out by $\mathbb{Z}_2$ under which $x\rightarrow -x$, in the case of an interval 
compactification, we must only consider the combinations $\frac{1}{2}\left(e^{inx}+e^{-inx}\right)$ in the 
KK modes.

Turning next to the dimensional reduction of a free boson and a free fermion, we start with the action
\be
   S=\int dt\int dx \left[-\frac{1}{2}\left(\d_\mu\widetilde \phi\right)^2-\overline{\widetilde\psi}\dslash\widetilde\psi\right]\,.
\ee
Taking as an ansatz $\phi=\phi(t),\psi=\psi(t)$, using $\int dx=2\pi R$, and rescaling
$\phi=\sqrt{2\pi R}\tilde\phi$, $\psi=\sqrt{2\pi R}\tilde \psi$ gives
\be
   S_{\rm 1d}=\int dt \left[\frac{\dot \phi^2}{2}+\psi_1^\dagger \dot \psi_1
   +\psi_2^\dagger \dot\psi_2\right].
\ee
However, as we have already argued, since the bosonization mixes kinetic and interaction terms in the 
action, it can in principle mix zero with nonzero modes. If in addition we rescale the coupling as
\be
   \b =\frac{\tilde \b}{\sqrt{2\pi R}}\,,
\ee
the dimensionally reduced bosonization formulae in (1+1)-dimensions read 
\bea
\psi_{1,n}&=&C :e^{-i\frac{2\pi}{\b}n\epsilon\dot \phi-i\frac{\b}{2}\phi}:\cr
\psi_{2,n}&=&-iC :e^{-i\frac{2\pi}{\b}n\epsilon\dot \phi+i\frac{\b}{2}\phi}:\,,\label{bose1}
\eea
where now $C$ is dimensionless and incorporates the $(\mu R)$ factor. We note at this juncture that, since the exponent contains a factor 
of the length of the spatial circle, it is not possible to fully dimensionally reduce on the fermionic side. Still, let us proceed and see what 
can be learnt. For future purposes, we will discretize this factor in the exponential as $n\epsilon$. The reduction of the duality relation between currents reduces to $j_0=0$ and
\be
j_1=j=-\frac{i\b}{2\pi}\dot\phi\;.
\ee
On the other hand, $j\equiv \psi^\dagger_{1,n+1}\psi_{1,n}-\psi^\dagger_{2,n+1}\psi_{2,n}$. Putting this together then, the scalar kinetic 
term would come from a 4-fermi (current-current) interaction, as before, since
\be
-\frac{g}{2}\left( \psi^\dagger_{1,n+1}\psi_{1,n}-\psi^\dagger_{2,n+1}\psi_{2,n}\right)^2\equiv
-\frac{gj^2}{2}=\frac{g\b^2}{8\pi^2}\dot \phi^2.\label{currentint}
\ee
Alternatively, if we simply (and naively) remove the $n\epsilon$ from the exponent giving
\bea
\psi_{1}&=&C :e^{-i\frac{2\pi}{\b}\dot \phi-i\frac{\b}{2}\phi}:\cr
\psi_{2}&=&-iC :e^{-i\frac{2\pi}{\b}\dot \phi+i\frac{\b}{2}\phi}:\,,\label{bose2}
\eea
for the bosonization formulae, we would not be able to obtain the kinetic term as above. In both (\ref{bose1}) and (\ref{bose2}) cases above though, we find
\bea
\psi_{1,n}^\dagger \psi_{2,n}\sim \psi_1^\dagger\psi_2&\sim&  :e^{i\b \phi}:\cr
\psi_{2,n}^\dagger \psi_{1,n}\sim \psi_2^\dagger \psi_1&\sim& :e^{-i\b\phi}:\,,\label{invbose1}
\eea
so that, as in (1+1)-dimensions, the fermion mass term produces a cosine interaction for the boson,
\be
\psi^\dagger_1\psi_2+\psi^\dagger_2\psi_1\sim :\cos (\b \phi) :
\ee
That means that the massive fermions, $\psi_n$, with current-current interaction (\ref{currentint}) are equivalent to a sine-Gordon boson in one 
dimension, with kinetic term. However, in the case of (\ref{bose2}), we cannot obtain a current-current interaction which arises from the scalar 
kinetic term and depends crucially on having a space variable $x$ to integrate over, as well as being able to separate $x$ from 
$y$.\footnote{In the case of (\ref{bose1}) we discretized the spatial interval, but the effect was the same} This is no longer possible in one dimension.

\subsection{Luttinger liquid bosonization and the SYK model}
To see how to treat current-current type interactions appropriate for the ($q=4$) SYK model, let us look at the second possibility, that of increasing 
the number of fermionic degrees of freedom. One way to accomplish this is to introduce a large set of KK modes from the circle expansion of 
the (1+1)-dimensional model. Another, is to consider some large number, $2N$ say, of massive fermions $\psi_{1,n}$ and $\psi_{2,n}$,  for 
$n=1,...,N$,  with current-current (4-fermi) interaction (\ref{currentint}) and bosonization (\ref{bose1}). This corresponds to a dimensional 
reduction/discretization of the 1+1 dimensional bosonization relation. This 4-fermi interaction (\ref{currentint}) is of the same form of the 
generic SYK model coupling $J_{ijkl}\psi^i\psi^j\psi^k\psi^l$, if we consider the formal replacement of the Majorana fields
\be
   \psi^i\rightarrow(\psi_{1,n},\psi^\dagger_{1,n},\psi_{2,n},\psi_{2,n}^\dagger).\label{SYK1}
\ee
Of course, there is no guarantee that, with $J_{ijkl}$ random but restricted to only nonzero terms that give the interactions (\ref{currentint}) 
as well as the constraints that come from the Majorana condition, we have the same remarkable behaviour exhibited by the original SYK 
model. However, the bosonized model has a kinetic term, which we can take to mean that {\em the dualization (\ref{bose1}) is valid for 
any value of the coefficient of the cosine potential for the boson}. This needs to be studied further. 

Of course, we could also consider the case when the potential for the boson is very large relative to its kinetic term. Similarly, we would 
like to be able to ignore also the fermion kinetic and mass terms, as is encountered in the usual SYK model. This would lead us to the 
bosonization rules (\ref{bose2}) and, subsequently, to the possibility of having much fewer fermions, as there is no need for an index 
$n$ now. But, as we saw earlier, we would still need to obtain a problematic 4-fermi-type interaction.

The solution to our conundrum lies in a more minimalist approach, in which we double the number of fermions and mimick the 
bosonization of a Luttinger liquid (see for instance chapter 5 of \cite{Nastase:2017cxp} for a review). Recall that there one starts 
from the (fermionic) Hubbard model for a one-dimensional gas of electrons. Linearizing the system near the fermi surface and 
taking a continuum limit, yields a one-dimensional model with massive 4-fermi interaction which is subsequently solved by 
bosonization. The left- and right-moving modes inherit the spin degrees of freedom from the electrons on the lattice so that 
when the system is diagonalized (via a Bogoliubov transformation), it separates into a ``spinon" mode which carries spin but 
no charge, described by the sine-Gordon model, and a charged but spinless ``holon" degree of freedom which is free.\\

\noindent
Based on this Luttinger liquid formulation of bosonization, we now sketch how this simpler version of 1-dimensional bosonisation 
works. We start by renaming $(\psi_1,\psi_2)$ as $(\psi_+,\psi_-)$. To these fields we add another binary (spin) index 
$\sigma=\uparrow,\downarrow$. Using this, we can then write down two copies of the bosonization relation (\ref{bose2}) as, 
\be
\psi_{\pm,\sigma}=C_\pm : \exp\left[-i\left(\frac{2\pi}{\b}
\dot\phi_\sigma\pm \frac{\b}{2}\phi_\sigma\right)\right]:\;.
\ee
This in turn gives two copies of the inverse relation (\ref{invbose1}), 
\bea
\psi^\dagger_{\pm \uparrow}\psi_{\mp \uparrow}&\sim& :e^{\pm i \b \phi_\uparrow}:\cr
\psi^\dagger_{\pm \downarrow}\psi_{\mp \downarrow}&\sim& :e^{\pm i\b \phi_\downarrow}:\;,
\eea
with the resulting  fermion bilinears combining into the mass terms,
\be
\psi^\dagger_{+\sigma} \psi_{-\sigma}+\psi^\dagger_{-\sigma}\psi_{+\sigma}\sim :\cos (\b \phi_\sigma) :.\label{massterm}
\ee
Now, however, we can construct a different type of 4-fermi term (one appropriate for a 
Luttinger liquid construction),  
\be
:\psi^\dagger_{+\uparrow}\psi_{-\uparrow}: :\psi^\dagger_{-\downarrow}\psi_{+\downarrow}:+h.c.
\sim :\cos(\b(\phi_\downarrow-\phi_\uparrow)):. \label{special4fermi}
\ee
To summarize; a 4-fermi term of the form (\ref{special4fermi}), with or without the mass term (\ref{massterm}) for the 
fermions will map under the duality to a bosonic model for the scalars $\phi_\uparrow, \phi_\downarrow$ with cosine 
potential (\ref{special4fermi}) and a possible mass term (\ref{massterm}), but {\it without a kinetic term}.

In this sense, this construction yields a bosonization in (0+1)-dimensions that maps a (non-kinetic) doubled sine-Gordon 
model to a Thirring-like model with the above 4-fermi interaction. Having thus treated a 4-fermi interaction means that we 
can now again contemplate constructing an SYK-like model with a bosonic counterpart. Of course, now, we would need to 
consider $N$ copies of the model (for $n=1,...,N$) with different couplings for the fermions, drawn at random from some 
distribution. We would expect the corresponding bosonized theory to consist of $N$ copies of a sine-Gordon-like model with correspondingly random couplings.

More precisely, we would expect an SYK-like model with couplings $J_{ijkl}$, again drawn at random from some Gaussian distribution, but where we replace each latin index, say $i$ by $(n,\pm, \sigma)$ (and with $n=1,...,N$). In this set, the only nonzero couplings are those with 
\be
(ijkl)\rightarrow (+\uparrow n, -\uparrow n, -\downarrow n,+\downarrow n).\label{SYK2}
\ee
This replacement should not be thought of as  ``restricting the indices'', either in (\ref{SYK1}) or in (\ref{SYK2}), but rather  as ``gaining indices'' with respect to the usual SYK model,  depending on whether we think of the index $i$ of SYK as corresponding to $(n,\pm, \sigma)$, or to be included in $n$ as $n=(ijkl)$. Either way, the bosonization leads to the doubled sine-Gordon scalar model above, in 0+1 dimensions, with random $\b_i$. Importantly, the physics is not guaranteed to match the original SYK model and is something that would require explicit checking. 

\section{Action on SYK-like models and gravity dual}
The utility of bosonization as a field theory duality is two-fold: (i) typically, it maps an interacting fermion problem to a free 
boson\footnote{Of course, with the usual caveats that, for example, the fermion mass term gets mapped to a non-trivial cosine-type 
interaction, as we have already seen in the context of the massive Thirring/sine-Gordon correspondence.}, and (ii) it is a nonperturbative 
method which maps the strong (weak) coupling regime of the fermionic theory into the corresponding weakly (strongly) coupled bosonic 
theory. This latter aspect is particularly interesting if the strongly coupled field theory in question possesses a (weakly coupled) gravity dual, 
since it hints at the intriguing possibility of a composite duality between a {\it weakly coupled} quantum theory and a {\it weakly coupled} 
classical gravity theory. In the previous section, we gave a sketch of how to go about bosonizing an SYK-like model in $(0+1)$-dimensions. 
Part of our motivation for attempting this in the first place was precisely because the SYK model is believed to have a gravity dual. In order 
to explore this further, let us ask how the bosonization `acts' on the gravity side, begining with a review of some of the salient features of the 
proposed gravity dual of the SYK model. 

\subsection{Jackiw-Teitelboim gravity from near-horizon extremal BTZ}

In \cite{Das:2017pif,Das:2017wae} further evidence was presented in support of the  conjecture that the gravity dual of the SYK model 
is the Jackiw-Teitelboim theory of two-dimensional dilaton gravity \cite{Maldacena:2016hyu}, whose dynamics is captured by action
\be
S=-\frac{1}{16\pi G}\int d^2x \sqrt{-g}[\phi(R+2)-2\phi_0]\;,
\ee
and which possesses a $T=0$, $AdS_2$ and dilaton background solution
\bea
ds^2&=& \frac{-dt^2+dz^2}{z^2}\cr
\phi(z)&=& \phi_0+a/z\;.\label{background}
\eea
This in turn can be interpreted as an $AdS_2\times S^1/\mathbb{Z}_2$ type geometry with metric
\be
ds^2=\frac{-dt^2+dz^2}{z^2}+\left(1+\frac{a}{z}\right)^2dy^2\,,\label{ads3}
\ee
arising from the near horizon limit of a charged extremal BTZ black hole in three dimensions. Here, since $S^{1}/\mathbb{Z}_{2}$ is topologically 
an interval, we will take $-L<y<L$. Note that at the $z=0$ boundary of $AdS_2$, the metric of the $S^1/\mathbb{Z}_2$ also becomes infinite, so 
that for some small $\epsilon$ and $z=\epsilon$,
\be
ds^2\simeq \frac{-dt^2+a^2dy^2}{z^2}\;,\label{boundary}
\ee
giving a boundary (from the Penrose diagram construction) that is topologically $\mathbb{R}_t\times S^1/\mathbb{Z}_2$. To understand 
how \eqref{background} arises as a near-horizon limit of an extremal 3-dimensional black hole, we can start from a $J=0$ BTZ 
solution \cite{Carlip:1995qv}, for which
\be
   ds^2=-\left(\frac{r^2}{l^2}-8GM\right)dt^2+\frac{dr^2}{\frac{r^2}{l^2}-8GM}+r^2d\phi^2\;.
\ee
As usual in 3-dimensional gravity, $M\geq 0$ produces a black hole background, while $M=-1$ corresponds to $AdS_3$ space. For more general 
values of $M<0$, the geometry has a conical deficit $\Delta \phi=2\pi \a$, related to $M$ by $M=-(1-\a)^2$. In particular then, the pure $AdS$ 
geometry has no conical deficit. \\

\noindent
For nonzero angular momentum  $J$, the BTZ solution takes the form 
\be
ds^2=-f^2(r)dt^2+\frac{dr^2}{f^{2}(r)}+r^2(Ndt +d\phi)^2\;,
\ee
with
\be
f^2(r)=-8GM+\frac{r^2}{l^2}+\frac{J^2}{4r^2}\;,\;\;\;
N=-\frac{J}{2r^2}.
\ee
In this case, when $M\geq |J|>0$, the background corresponds to a black hole while, for $M<|J|$, and 
$M\neq -1$, it possesses a naked singularity. Let us focus on the black hole. In this case, there are two 
(an inner and an outer) horizons, at 
\be
r_\pm^2=\frac{Ml^2}{2}\left[1\pm \left(1-\frac{J}{Ml}\right)^2\right]\;,
\ee
and which merge as $J\to0$. There is also a coordinate transformation (see eq. 2.9 in \cite{Carlip:1995qv})
for $r>r_+$ that turns $(t,r,\phi)$ into $(x,y,z)$, and results in the $AdS_3$ metric,
\be
ds^2=l^2\frac{dx^2-dy^2+dz^2}{z^2}.
\ee
This transformation is singular in the $J\rightarrow 0$ limit, since then $z=z(r,t,\phi)\rightarrow 0$.
A charged BTZ black hole \cite{Martinez:1999qi} is as above with $N=0$. The metric for this geometry is given by
\be
ds^2=-f^2(r)dt^2+\frac{dr^2}{f^{2}(r)}+r^2d\phi^2\;,
\ee
with 
\bea
f^2(r)=-8GM+\frac{r^2}{l^2}-8\pi G Q^2\ln \frac{r}{l}\,,\qquad
F_{tr}=\frac{Q}{r}.
\eea
Again, a change of coordinates (see Appendix A of \cite{Maity:2009zz}) brings this into the form 
\bea
ds^2&=&\frac{1}{z^2}\left[-f^2(z)dt^2+\frac{dz^2}{f(z)}+dx^2\right]\cr
f(z)&=&1-z^2+\frac{Q^2}{2}z^2\ln z\,,
\eea
with vector potential $A= Q\ln z\, dt$. In this form it is clear that the geometry has a boundary at $z=0$, horizon at $z=1$, and a temperature 
\be
T=\frac{f'(1)}{4\pi}=\frac{1-Q^2/4}{2\pi}\;.
\ee
The extremal case can then be read off as corresponding to a charge satisfying $Q^2=4$. 
Expanding the solution around the horizon at $z=1$ in this extremal case, we get
\be
f^2(z)\simeq 2(1-z)^2\equiv 2 \tilde z^2\;,\;\;\;
A\simeq -2(1-z)dt\simeq -2\tilde z dt.
\ee
giving a metric 
\be
ds^2\simeq -2\tilde z^2 dt^2+\frac{d\tilde z^2}{2\tilde z^2}+dx^2\;.
\ee
This, however, is not quite the same as \eqref{background} so, evidently, something else is required. In fact, the solution (\ref{background}), including the 
dimensional reduction of the 3 dimensional gravity, was found in \cite{Cadoni:2008mw} (see also \cite{Almheiri:2014cka} and \cite{Maldacena:2016upp}, 
for more relevant details and relation to the SYK model). The difference now, is in {\em how one takes the near-horizon limit}, in combination with the 
extremal limit\footnote{Note however that in \cite{Almheiri:2014cka}, the authors take 2-dimensional dilaton gravity theories to have action 
\be
S=\frac{1}{16\pi G}\int d^2x\sqrt{-g}[\phi^2R+\lambda(\d_\mu\phi)^2-U(\phi)]\;,\nonumber
\ee and point out that by Weyl rescalings of the metric one can remove the kinetic term for the scalar, so one can put $\lambda=0$ in the 
general form above without loss of generality.}. Dimensional reduction of the 3-dimensional gravity action is implemented on the angle $\phi$, by the ansatz
\be
ds^2_3=ds_2^2+l^2\phi^2d\theta^2\;,
\ee
with $\phi$ now taken to be a dilaton. A subsequent rescaling by $\eta=(l/4G)\phi$, gives the action of a Jackiw-Teitelboim gravity coupled to a scalar and Maxwell field,
\be
S=\frac{1}{2}\int d^2x\sqrt{-g}\eta\left(R+\frac{2}{l^2}-4\pi G F^2\right)\;.
\ee
This theory has a solution with linear dilaton, and 
\be
ds^2=-f^2(r)dt^2+\frac{dr^2}{f^2(r)}\;,\;\; F_{\mu\nu}=\frac{Q}{r}\epsilon_{\mu\nu}\;,\;\; \eta=\eta_0\frac{r}{l}.
\ee
On this solution, the $F^{2}$ term in the action becomes a constant, as considered in \cite{Das:2017wae}.

\subsection{The action of bosonization on the gravity dual}

Let us focus now on the gravity dual in the 3-dimensional form (\ref{ads3}), with near horizon boundary (\ref{boundary}). The natural boundary theory 
is an infinite set of quantum mechanical modes, arising from the KK expansion on the finite line $S^1/\mathbb{Z}_2$, parametrized by $y$. But this 
is slightly different from the analysis of \cite{Das:2017wae}, allowing for the possibility of two different formulations for the boundary theory related by 
bosonization. Note that now we obtain boundary theories living in 1+1 dimensions, but KK expanded on $S^1/\mathbb{Z}_2$, so according to our 
analysis, bosonization is guaranteed to work for them. It is not completely clear to us if the bosonization will continue to hold in the case of a finite 
number of modes in 0+1 dimensions. 

In order to interpret the action of the boundary bosonization on the full gravity dual, it will be useful to embed the bulk geometry in string theory. 
Our strategy will be to interpret the Jackiw-Teitelboim gravity as arising from string theory via dimensional reduction in order to relate its scalar 
$\phi$ to the 10-dimensional string theory dilaton, $\varphi$. We start from the 10-dimensional action  
\be
S=\frac{1}{2\kappa_{10}^2}\int d^{10}x\sqrt{-g^{(10)}}\,e^{-2\varphi}\left[{\cal R}^{(10)}+4(\d_\mu\varphi)^2+...\right].
\ee
A simple toroidal reduction on a constant compact space, reduces this to 
\be
S=\frac{1}{2\kappa^2_2}\int d^2x\sqrt{-g^{(2)}}\,e^{-2\varphi}\left[{\cal R}^{(2)}+4(\d_\mu\varphi)^2+...\right].
\ee
Then, a general Weyl rescaling $g_{\mu\nu}=e^{\a\varphi}\tilde g_{\mu\nu}\equiv \Omega^2\tilde g_{\mu\nu}$ applied to the 2-dimensional action gives 
\bea
S&=&\int d^2x\sqrt{-\tilde g^{(2)}}\,e^{-2\varphi}\left[\tilde {\cal R}^{(2)}+4(\d_\mu\varphi)(\d_\nu\varphi)\tilde g^{\mu\nu}
+2\a\tilde g^{\mu\nu}\tilde\nabla_\mu\tilde \nabla_\nu \varphi+...\right]\cr
&=&\int d^2x\sqrt{-\tilde g^{(2)}}\,e^{-2\varphi}\left[\tilde {\cal R}+(4+4\a)(\d_\mu\varphi)(\d_\nu\varphi)\tilde g^{\mu\nu}+...\right].
\eea
This means that we can tune the coefficient of the $\varphi$ kinetic term by a Weyl rescaling. In particular we can set it to zero by 
choosing $\a=-1$, so that $\phi=e^{-2\varphi}$. Having thus related the solution to 10-dimensional string theory, allows us to take 
advantage of the symmetries of the latter. Specifically, we note that the solution is invariant under a T-duality in time (similar to that 
in the scattering amplitude case of Alday and Maldacena in \cite{Alday:2007hr}), coupled with renaming $x=1/z$. Together, these change the dilaton to 
\be
\tilde \varphi=\varphi-\frac{1}{2}\ln g_{00}\Rightarrow \tilde \phi=\phi g_{00}.
\ee
To fix this, note that near $z= 0$, $\phi\simeq a/z$ and recall that under an S-duality $\phi\to1/\phi$). Then under a sequence of STS dualities, we note that
\be
\phi=\frac{a}{z}\stackrel{\mathrm{S}}{\longrightarrow} \frac{z}{a}\stackrel{\mathrm{T}}{\longrightarrow} 
\frac{1}{z^2}\frac{z}{a}=\frac{1}{az}=\frac{x}{a}\stackrel{\mathrm{S}}{\longrightarrow} \frac{a}{x}\;.
\ee
In other words, the STS transformation leaves the gravity dual invariant {\em in the UV region (near the boundary)}. We observe that 
the STS transformation maps the fields to (approximately) themselves (and, consequently, the worldsheet theory to itself also) after 
the redefinition $\widetilde{X}^0=R/X^0$. From the point of view of holography, this amounts to a UV$\longleftrightarrow$IR reversal. 

We should check then that bosonization shares this property. For concreteness, let us consider the sine-Gordon/massive Thirring model duality. 
In the bosonic theory, since $\a= m^2$ and $\a\b^2=\lambda$ , large $\b^2$ corresponds to small $g$. Consequently, at high energies where 
$\a\ll E^2$  and the kinetic term dominates over the potential, $\b \ll 1$. This translates into $g^2\gg 1$ in the massive Thirring model. However, 
there high energy means $E\gg m, j_\mu$, which translates into small $g$, so that the kinetic term dominates over the potential. We see that 
this corresponds to the opposite regime from the the sine Gordon model, so indeed, high energy is mapped to low energy. Another way of 
understanding this UV/IR reversal is that the low energy limit of the sine-Gordon model means, at least in some region of the parameter 
space of the theory where the soliton mass ($\sim m^3/\lambda\sim\sqrt{\a}/\b^2 $) is much less than the fundamental mass, i.e., when 
$\b\ll 1$,  that we need to ignore the solitons, and focus on the free theory, thus on the kinetic term. However, since the soliton of the 
sine-Gordon model maps to the fundamental (fermionic) particle of the massive Thirring model, ignoring it means ignoring the kinetic 
term of the massive Thirring model, which is a high energy limit. Finally, note that since $\psi\sim \exp\left[-i \int^x dx' \dot\phi(x')-i\b 
\phi(x)\right]$ for $\phi\sim e^{iEt}$ the contributions of the large $E$ modes to the behaviour of $\psi$ at high energies, also 
consistent with the previous observations.

\section{Conclusions}

This paper details our proposal for a bosonization duality in one spacetime dimension. Our strategy was to begin by KK-reducing 
the familiar 2-dimensional bose-fermi duality. We showed that a naive dimensional reduction is, however, deficient since, with no 
space variable to integrate over, we are unable to obtain any current-current interaction terms. We then argued that this could be 
circumvented by adding more fermions and implementing an analog of Luttinger liquid bosonization. We found this more promising
in that it has the potential to bosonize quantum mechanical systems with 4-fermi interactions and hence also the $q=4$ SYK model. 
Of course, since the SYK model is fundamentally random, we would still need to understand how such a bosonization acts on 
quenched random couplings. This is nontrivial but, in principle at least, treatable with the replica method. One reason to focus on 
the SYK model is that with a known bulk dual, it gives us the opportunity to formulate a holographic dual to bosonization.  We took
the first steps in this direction by sketching how to embed the JT gravity dual to the SYK model in string theory. The action of the 
boundary bosonization then corresponds to a sequence of S and T dualities that leave the background almost invariant (near the 
boundary) with a commensurate exchange of low and high energies.

Clearly there is much that still needs to be done, both in testing the duality, and its proposed action on the gravity dual. We hope that
the interest in bosonizations in two and three dimensions will translate into interest in a 1-dimensional version, and that there is as 
much to learn from it as its higher-dimensional avatars.

\section*{Acknowledgements}
We would like to thank Micha Berkooz, Juan Maldacena, Joan Simon, Spenta Wadia and Tony Zee for useful discussions. JM is supported in 
part by the NRF of South Africa under grant CSUR 114599 and the National Science Foundation under Grant No. NSF PHY-1748958. 
JM would like to thank the organisers and participants of the ``Chaos and Order 2018" program at the KITP of the University of California, 
Santa Barbara for a stimulating and productive environment during the final stages of this work. The work of HN is supported in part by CNPq grant 304006/2016-5 and 
FAPESP grant 2014/18634-9. HN would also like to thank the ICTP-SAIFR for their support through 
FAPESP grant 2016/01343-7, and to the University of Cape Town for hospitality during the final stages of this
work. 
\bibliography{DualityOneD}
\bibliographystyle{utphys}

\end{document}